# Empirical Analysis of Service Quality, Reliability and End-User Satisfaction on Electronic Banking in Nigeria

Esther Enoch Yusuf[1], Abubakar Bala[2]

[1](Department of Banking and Finance, School of Business and Management Technology / Federal Polytechnic, Mubi, Nigeria)

[2](Department of Business Adm., Faculty of Art & Social Science/ Gombe State University, Nigeria)

***Abstract:*** *Today, almost all banks have adopted ICT as a means of enhancing their banking service quality. These banks provide ICT based electronic service which is also called electronic banking, internet banking or online banking etc to their customers. Despite the increasing adoption of electronic banking and it relevance towards end users satisfaction, few investigations has been conducted on factors that enhanced end users satisfaction perception. In this research, an empirical analysis has been conducted on factors that influence electronic banking user's satisfaction perception and the relationship between these factors and the customer's satisfaction. The study will help bank industries in improving the level of their customer's satisfaction and increase the bond between a bank and its customer.*

## I. Introduction

The growing trend in the world of Information and Communication Technology has drastically changed the perspective of both bank customers and the banking industries on general banking activities. The combination of information technologies with functions of banks and financial institutions is called electronic banking. Electronic banking technologies have led banks and financial institutions to improve effectiveness of distribution channels through reducing the transaction cost and increasing the speed of service [1] [2].

Majority of banks in the developed world and some in the developing world are now offering electronic banking services with various degrees of complexities [3]. This gives the indication that e-banking is gradually taking the place of traditional banking services. In developing countries, for example, some banks have adopted e-banking as a way of communicating to customers with regards to issues concerning bank statements whiles other banks use internet banking services to allow customers to access their bank accounts and perform other banking transactions [4].

Many Banks have implemented full electronic banking services however; its adoption by end-users in Nigeria has been very slow even among the elites. This could be as results of unawareness or cyberphobia. For Banks to meet its aim of customer's satisfaction through electronic banking, then they must identify how the service is currently being perceived by adopters and potential adopters and the characteristics of such adopters. It is also the responsibility of the bank to determine whether there is a demand for such services and factors affecting the demand. It is against this that this study attempts to investigate and analyse the relationship between service qualities, reliability and end-user satisfaction on electronic banking.

In the next section, we present a review of related work on electronic banking and relationship between customers satisfaction and service quality. We then proposed a model to improve end-user adoption and satisfaction of electronic banking services. We also presented the research methodology and the research findings from the analysis of the empirical data and then conclude the paper.

## II. Related Work

**2.1    Concept of Electronic Banking**

The concept of electronic banking has been defined in ways; According to [5] Electronic Banking is an innovative service delivery mode that offers diversified financial services like cash withdrawal, funds transfer, cash deposits, payment of utility and credit card bills, cheque book requests, and other financial enquiries[5]. According to [6] electronic banking defined as any use of information and communication technology and electronic means by a bank to conduct transactions and have interaction with stakeholders.

The concept and scope of e-banking is still evolving. It facilitates an effective payment and accounting system thereby enhancing the speed of delivery of banking services considerably [7]. [8] Argues that electronic banking is a product of e-commerce in the field of banking and financial services. In what can be describe as business to consumer domain for balance enquiry request for cheque books recording stop payment instruction balance transfer instruction account opening and other forms of traditional banking service. Electronic Banking can also be defined as depicted below:

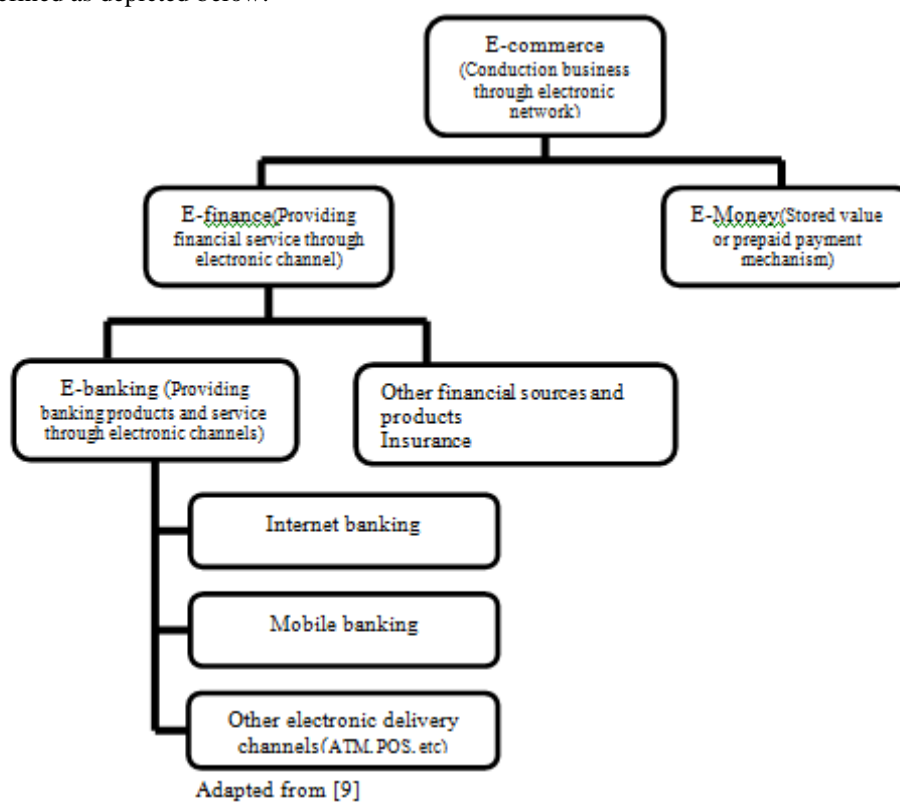

Adapted from [9]

Electronic banking has been around for quite some time in the form of automatic teller machines (ATMs) and telephone transactions. In more recent times, it has been transformed by the internet – a new delivery channel that has facilitated banking transactions for both customers and banks [10].

**2.2 Service Quality, Reliability and Customers Satisfaction**
Today, almost all banks have adopted ICT as a mean of enhancing service quality of banking services. This banks provide ICT based electronic services to their customers which is also called electronic banking, internet banking or online banking etc. It brings involvement, customer centricity, enhance service quality and cost effectiveness in the banking services and increasing customers' satisfaction in banking services [11].

Customer satisfaction can be defined as a "disconfirmation paradigm" since it is a result of confirmation/disconfirmation of expectation that evaluates a product's performance with it expectation and desire [12]. Customer satisfaction is therefore an attitude or a rating made by the customer by comparing their pre-purchase expectation to their subjective perceptions of actual performance [13]. ''Satisfaction is a person's feeling of pleasure or disappointment resulting from comparing a product's performance (outcome) in relation to his or her expectation'' [14].

Bank customer satisfaction is regarded as banks fully meeting the customers' expectation [15] and also said to be a feeling or attitude formed by bank customers after service, which expressly connects the various purchasing behaviour [16].

[17] [18] have strongly established Service quality as a predictor of customer satisfaction particularly, in electronic services. When perceived service quality is less than expected service quality customer will be dissatisfied [19]. According to [20] satisfaction super ordinate to quality-that quality is one of the service dimensions factored in to customer satisfaction judgment. The higher level of perceived service quality results in increased customer satisfaction. Additionally, [21] indicated that dissatisfaction with the electronic or internet banking is because of the high failure rates of most of the innovative products and services introduced.

Reliability defines the level of transaction security, promptness and focus on the elements that may contribute to user trust. According to [22], success of electronic banking depends on electronic banking service and reliability. Reliability is established in some studies as a key factor that most customers consider before and even during usage of electronic banking service [23]. Prior researches have revealed that reliable/prompt responses, attentiveness, and error-free electronic banking platforms have a considerable impact on customer satisfaction [23] [4].

**2.3　Research Objectives**
The main objective for conducting this research is to determine the impact of electronic banking services delivery on end user satisfaction in Nigeria. The specific objectives to be considered are as follows:
  i. To assess electronic banking service availability, convenience, efficiency, fulfillment, privacy and reliability in the Nigerian banking industry.
  ii. To analyse the relationship between service qualities and end-user satisfaction on electronic banking.

**2.4　Hypotheses**
This study proposes the following hypothesis:
$H_1$: $X_k$ does not have positive impact on customer satisfaction.
$H_2$: Network quality positively affects end-user service quality perception.
$H_3$: A higher perception of service quality positively affects end-user satisfaction.
Where k =1, 2, 3, 4 ,5 and 6 and $X_1$, $X_2$, $X_3$, $X_4$, $X_5$ and $X_6$ represent electronic banking service availability, convenience, efficiency, fulfillment, privacy and reliability respectively.

### III.　Conceptual Framework

This research offered a conceptual framework on how electronic banking availability, quality, convenience and reliability positively affects customer satisfaction based on [17] [18] [20] [25] [26] [27] [28] [29] [30] [31] [32]. The conceptual framework and relationships between electronic banking availability, quality, convenience and customer satisfaction is depicted in Fig. below:

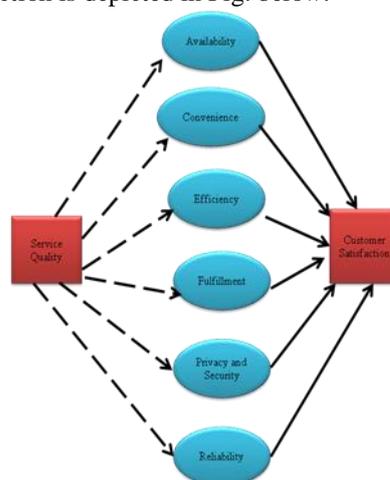

Research Conceptual framework

### IV.　Research Methodology

The conceptual framework was developed on the basis of the literature review. Keeping in view the conceptual framework, the study requires the collection of primary and secondary data both.

A questionnaire is developed with three main variables that are end user satisfaction, electronic banking quality and service reliability. The questionnaire was self-administered by the researcher.

SPSS 20.0 was used for testing questionnaire reliability and it addition, it is also used for the descriptive and correlation analysis of the data collected.

The questionnaire was divided into two sections. Section 1 is the demographic study and it is made up of questions pertaining the respondents' demographic profiles, such as age, gender, marital status, educational qualification, occupation, employment sector and designation. While the section 2 consists of majorly the end user opinion on electronic banking quality, service reliability and end user satisfaction. All the questions of the questionnaire from section 2 used a Likert scale ranging from 1 = Strongly Agree to 5 = Strongly Disagree.

The entire population of end users that adopt electronic banking service in Adamawa and Gombe State are used as the universe for this study. We then adopted a convenience sampling on electronic banking users from the selected states' banks within the state capital to conduct our study.

## 4.1 Result, Analysis and Discussion
**Demographic**

Table I: Demographic Profile of respondents

| Demographic Variable | Frequency | Percentage |
|---|---|---|
| **Age** | | |
| 18 - 30 years | 82 | 22.2 |
| 31 - 45 years | 150 | 40.7 |
| 46 - 60 years | 110 | 29.8 |
| above 60 years | 27 | 7.3 |
| Total | 369 | 100.0 |
| **Gender** | | |
| Male | 202 | 54.7 |
| Female | 167 | 45.3 |
| Total | 369 | 100.0 |
| **Marital Status** | | |
| Single | 227 | 61.5 |
| Married | 138 | 37.4 |
| Divorced/Seperated | 4 | 1.1 |
| Total | 369 | 100.0 |
| **Qualification** | | |
| Bachelors | 235 | 63.7 |
| MSc/PhD | 86 | 23.3 |
| SSCE/Dip | 48 | 13.0 |
| Total | 369 | 100.0 |
| **Number of Banks presently Using** | | |
| 1 | 116 | 31.4 |
| 2 | 199 | 53.9 |
| 3 | 49 | 13.3 |
| 4 | 4 | 1.1 |
| 5 | 1 | .3 |
| Total | 369 | 100.0 |
| **Period of Electronic Banking Usage** | | |
| Below 6 months | 37 | 10.0 |
| Between 6 months to 1 yr | 215 | 58.3 |
| above 1 yr | 117 | 31.7 |
| 4 | 1 | .3 |
| Total | 369 | 100.0 |

Source: field study

A total of 500 questionnaires were administered; 250 were administered in Adamawa State and the other 250 in Gombe State From the total of 500 questionnaires distributed, 369 were fully completed. The gender of the respondents was well represented, males accounted for 54.7% of total respondents whilst females accounted for 45.3%.

The age distribution of respondents ranged between 18 to 70 years of age, with the largest group of respondents aged between 18 and 45 years old. From the users 61% were single. Among the respondents 13% had SSCE/Diploma, 63.7% had Bachelors degree and 23.3% had Masters/Ph.D.

From the demographic section, it is evident that 62.9% of the respondents were young (below 45 years), 37.1% were above the age of 45.

## 4.2 Test of Hypothesis
This part of the study tested the various hypotheses of this study. In testing the hypotheses Spearman correlation coefficient was used.
$H_1: X_k$ does not have positive impact on customer satisfaction

*Empirical Analysis of Service Quality, Reliability and End-User Satisfaction on Electronic Banking…*Where k =1, 2, 3, 4 ,5 and 6 and $X_1$, $X_2$, $X_3$, $X_4$, $X_5$ and $X_6$ represent electronic banking service availability, convenience, efficiency, fulfillment, privacy and reliability respectively.

The result of the Spearman correlation is presented in table below:

Table II: relationship between availability, convenience, efficiency, fulfillment, privacy, reliability and end user Satisfaction

| | | | Fulfillment | Availability | Reliability | Convenience | Efficiency | Privacy |
|---|---|---|---|---|---|---|---|---|
| Spearman's rho | Fulfillment | Correlation Coefficient | 1.000 | .080 | 1.000** | .053 | .041 | .045 |
| | | Sig. (2-tailed) | . | .127 | . | .307 | .427 | .386 |
| | | N | 369 | 369 | 369 | 369 | 369 | 369 |
| | Availability | Correlation Coefficient | .080 | 1.000 | .080 | .029 | .056 | .535** |
| | | Sig. (2-tailed) | .127 | . | .127 | .585 | .281 | .000 |
| | | N | 369 | 369 | 369 | 369 | 369 | 369 |
| | Reliability | Correlation Coefficient | 1.000** | .080 | 1.000 | .053 | .041 | .045 |
| | | Sig. (2-tailed) | . | .127 | . | .307 | .427 | .386 |
| | | N | 369 | 369 | 369 | 369 | 369 | 369 |
| | Convenience | Correlation Coefficient | .053 | .029 | .053 | 1.000 | .645** | .005 |
| | | Sig. (2-tailed) | .307 | .585 | .307 | . | .000 | .922 |
| | | N | 369 | 369 | 369 | 369 | 369 | 369 |
| | Efficiency | Correlation Coefficient | .041 | .056 | .041 | .645** | 1.000 | .024 |
| | | Sig. (2-tailed) | .427 | .281 | .427 | .000 | . | .650 |
| | | N | 369 | 369 | 369 | 369 | 369 | 369 |
| | Privacy | Correlation Coefficient | .045 | .535** | .045 | .005 | .024 | 1.000 |
| | | Sig. (2-tailed) | .386 | .000 | .386 | .922 | .650 | . |
| | | N | 369 | 369 | 369 | 369 | 369 | 369 |

**. Correlation is significant at the 0.01 level (2-tailed).

The table above shows the Spearman rho relationship between electronic banking service availability, convenience, efficiency, fulfillment, privacy, reliability and end user Satisfaction. The correlation is statistically significant at 0.01 level (2-tailed). We therefore accept the hypothesis.

$H_2$: Network quality positively affects end-user service quality perception.

Table III: Network quality positively affects end-user service quality perception

| | | | The electronic banking services provided has met my expectation | The Service of electronic banking is always available to me 24 hours per | I can login and perform transaction on any of my electronic banking service quickly and easily | Electronic banking is faster than the manual ordinary banking service |
|---|---|---|---|---|---|---|
| Spearman's rho | The electronic banking services provided has met my expectation | Correlation Coefficient | 1.000 | .038 | .037 | .653** |
| | | Sig. (2-tailed) | . | .466 | .474 | .000 |
| | | N | 369 | 369 | 369 | 369 |
| | The Service of electronic | Correlation Coefficient | .038 | 1.000 | .080 | .056 |

DOI: 10.9790/487X-171022834  www.iosrjournals.org  32 | Page

| | | | | | |
|---|---|---|---|---|---|
| banking is always available to me 24 hours per | Sig. (2-tailed) | .466 | . | .127 | .281 |
| | N | 369 | 369 | 369 | 369 |
| I can login and perform transaction on any of my electronic banking service quickly and easily | Correlation Coefficient | .037 | .080 | 1.000 | .041 |
| | Sig. (2-tailed) | .474 | .127 | . | .427 |
| | N | 369 | 369 | 369 | 369 |
| Electronic banking is faster than the manual ordinary banking service | Correlation Coefficient | .653** | .056 | .041 | 1.000 |
| | Sig. (2-tailed) | .000 | .281 | .427 | . |
| | N | 369 | 369 | 369 | 369 |

**. Correlation is significant at the 0.01 level (2-tailed).

$H_3$: A higher perception of service quality positively affects end-user satisfaction.

Table IV: A higher perception of service quality positively affects end-user satisfaction

| | | | Higher perception of | End-user Satisfaction |
|---|---|---|---|---|
| Spearman's rho | Higher perception of Service Quality | Correlation Coefficient | 1.000 | .978** |
| | | Sig. (2-tailed) | . | .000 |
| | | N | 369 | 369 |
| | End-user Satisfaction | Correlation Coefficient | .978** | 1.000 |
| | | Sig. (2-tailed) | .000 | . |
| | | N | 369 | 369 |

**. Correlation is significant at the 0.01 level (2-tailed).

The results indicate that there is a strong and positive relationship between higher perception of service quality and end user satisfaction (0.978). The correlation is statistically significant at 0.01 level (2-tailed). We therefore accept the hypothesis.

## V. Conclusion

In this research, electronic banking service; availability, convenience, efficiency, fulfillment, privacy and reliability were mainly focused to determine it implication on the level of end user satisfactions perception of electronic banking. Accordingly, majority of all the respondents were young and educated and they fully understood the composite of electronic banking services. For hypothesis testing, the data was analysed using Statistical Packages for Social Science (SPSS 20.0) to test the hypotheses and find the results for this research.

Two of the hypotheses have high positive correlation between each variable; this means that there is a high positive relationship service quality, reliability and end user satisfaction in electronic banking. Industrial banks needs to enhanced all these factors to enhance their customers satisfaction level and increase the percentage of electronic banking adaptors in Nigeria since it is found easier, faster, secured, reliable and more convenient than the manual banking system.